# Comparative study of the Portevin-Le Chatelier effect in interstitial and substitutional alloy

A. Sarkar, A. Chatterjee, P. Barat[*] and P. Mukherjee

*Variable Energy Cyclotron Centre*

*1/AF Bidhan Nagar, Kolkata 700 064, India*

Abstract:
*Tensile tests were carried out by deforming polycrystalline samples of an interstitial alloy, low carbon steel at room temperature in a wide range of strain rates where the Portevin-Le Chatelier (PLC) effect was observed. The observed stress time series data were analyzed using the nonlinear dynamical methods. From the analyses, we could establish the presence of marginal deterministic chaos in the PLC effect of the low carbon steel. Moreover, we made a comparative study of the PLC effect of this interstitial alloy with the substitutional Al-Mg alloy, which shows that the dynamics of the PLC effect in the interstitial alloy is more complex compared to that of the substitutional alloy.*

The deformation characteristics of some dilute substitutional and interstitial alloys often exhibit serrations when deformed in a certain range of temperatures and strain rates. The appearance of the serrations in the stress-strain curves of these materials is often designated as the Portevin-Le Chatelier (PLC) effect [1]. The PLC effect is a paradigm of instabilities of plastic flow and plastic strain localizations [2]. It has continuously drawn attention due to its intriguing spatio-temporal dynamics. In this respect, this effect falls into the class of complex non-linear driven systems exhibiting intermittent relaxation sequences.

---

[*] Corresponding author, Email: pbarat@veccal.ernet.in, Phone: +91-33-23371230, Fax: +91-33-23346871

Due to the inherent nonlinearity and the presence of multiple timescales, nonlinear dynamics has usually found its way in the study of PLC effect. In the last two decades, many statistical and dynamical studies have been carried out on the PLC effect [3, 4]. These studies provided a great deal of intricate details of the underlying dynamics of the effect. G. Ananthakrishna et. al. [5, 6] proposed a theoretical model of dislocation dynamics for the PLC effect which in a certain range of parameters yielded chaotic solutions. They later found the signature of chaos by analyzing the stress time series data obtained during the PLC effect [7]. This generated a new ebullience in this field and several authors studied the PLC effect in the light of chaos [8, 9]. But, these nonlinear dynamical studies of the PLC effect were completely confined to the substitutional alloys like Al-Mg [8, 9], Cu-Al [7] etc. The observation of the PLC effect in dilute interstitial alloys like low carbon steel is very old [10], however, there is no report to prove the existence of the chaotic behavior in the stress-time data of the PLC effect for these alloys.

For the first time, we have tried to establish the exact chaotic nature of the PLC effect of the low carbon steel when deformed at an intermediate strain rate regime at room temperature and compared our results with the reported results of the substitutional Al-Mg alloy.

The general consensus explains that the physical origin of the PLC effect is the Dynamic Strain Ageing (DSA) [2]. The locking and unlocking of dislocations in the solute environment is the essence of the DSA. In this respect, the microscopic mechanism of the DSA in the interstitial alloy is quite different from the substitutional alloy. The locking of dislocations in the substitutional alloys is attained by the formation of the Cottrell atmosphere of the solute atoms around it [11]. This requires vacancy-assisted diffusion of solute atoms over long distances and takes much longer time. On the contrary, there are two different mechanisms playing role in the locking of dislocations in the interstitial alloy, low carbon steel. One of them is the usual formation of Cottrell atmosphere of the solute carbon atoms by long-range diffusion [12, 13]. It is experimentally verified that at room temperature the diffusion coefficient of the interstitial carbon atom ($10^{-21} m^2 s^{-1}$) in iron [14] is much higher than that of the magnesium ($10^{-26} m^2 s^{-1}$) in Al-Mg alloy [15]. Hence, the diffusion of carbon atoms in low carbon steel is much faster than magnesium in the Al-Mg alloy. Again, carbon atoms occupy the interstitial positions in low carbon steel. Thus they do not require any vacancy assisted diffusion. Moreover, there is another locking mechanism for this interstitial alloy, called Snoek ordering. The interstitial solute atom carbon occupies the octahedral sites in the b.c.c crystals and produces tetragonal distortions. At zero stress field, the carbon atoms will occupy the three interstitial sites corresponding to the three directions of tetragonality with equal probability. Due to the stress field of dislocation, a redistribution of the interstitial atoms on different octahedral sites will take place with an increase in population of the sites with lower energy. By such a process, the energy of the system is lowered and this leads to the locking of dislocations. This is called the Snoek effect [16]. The locking due to the Snoek effect is accomplished merely by atomic rearrangement between the neighboring octahedral sites within a short time and also explains the rapid strain aging phenomena in low carbon steel [17]. Thus the pinning barrier due to the Snoek effect, although high, is extremely narrow of only one atom thick. This difference in the mechanism of DSA is expected to reflect in the stress-time series data for these two types of alloys during the PLC effect. Hence, it is interesting to make a comparative study of the dynamics of the PLC effect in these two types of alloys by analyzing the stress time series data.

In this work, we report the results of the detailed nonlinear studies on the PLC effect observed in the polycrystalline low carbon steel (0.15%C, 0.33%Mn, 0.04%P, 0.05%S, 0.15%Si and rest Iron). We have carried out tensile testing at room temperature on the cylindrical samples with the gauge length and diameter of 25 and 5 mm respectively. Samples were tested in a servo controlled INSTRON (model 4482) machine at a range of strain rates between $6.03 \times 10^{-5}$ sec$^{-1}$ and $5.66 \times 10^{-4}$ sec$^{-1}$. The PLC effect was observed throughout this range of strain rates. The stress-time response was recorded electronically at a periodic time interval (20 Hz). The stress data taken for the analysis were corrected for the strain hardening drift. A typical segment of the drift corrected true stress fluctuation vs. time curve for the strain rate $5.66 \times 10^{-4}$ sec$^{-1}$ is shown in Fig. 1. Similar tests [18] were also carried out on the Al-2.5%Mg alloy samples. Few of the analyses, which have not yet been reported in the literature were carried out on the observed experimental stress time series data of the Al-2.5%Mg alloy and are reported in the subsequent section where the results of the PLC effect for the two different alloys have been compared.

The dynamical analysis of the PLC effect starts with the reconstruction of the attractor from the one dimensional stress-time series data $\{x_j, j= 1, 2,..., N\}$ on the basis of the embedding theorems [19,20]. The trajectory in the embedding space consists of vectors given by $X_i = \{x_i, x_{i+\tau}, x_{i+2\tau},....., x_{i+(m-1)\tau}\}$, where $m$ represents the embedding dimension and $\tau$ is the delay time. The value of $m$ is estimated by the false nearest neighbor method [21]. $\tau$ is estimated from the first zero crossing of the auto-correlation function [21]. The singular value decomposition (SVD) method [22] is used to portray the attractor by removing the random noise present in the experimental data. The existence of a positive Lyapunov exponent is the most reliable signature of chaotic dynamics [23]. The estimation of the largest Lyapunov exponent is performed with the method proposed by Wolf et. al. [24]. We have used the Grassberger-Procaccia algorithm [25] to determine the correlation dimension ν of the attractor. The correlation entropy, $K_2$ is calculated using the prescription described in the reference [26]. For a chaotic system, $K_2 > 0$. So the estimation of $K_2$ from experimental time series data also serves as a tool to detect the chaotic nature of the dynamics.

The experimental stress time series data of the polycrystalline low carbon steel are analyzed using the above-mentioned methods. The false nearest neighbor method yielded the embedding dimension ($m$) ~ 8 and the delay time ($\tau$) obtained from the autocorrelation function is ~ 4 time units for all the data sets obtained at different strain rates. Fig. 2 shows the singular value spectrum for the strain rate $5.66 \times 10^{-4}$ sec$^{-1}$. It is clear that the important dynamics is confined to a 7-8 dimensional subspace of the embedding space. Fig. 3 shows the typical projection of the reconstructed attractor onto the two-dimensional subspace for the strain rate $5.66 \times 10^{-4}$ sec$^{-1}$. The values of the Lyapunov exponent for $m$=8 were between 0.050±0.007 and 0.350±0.026 for all the test data for various strain rates. This shows a marginal chaotic dynamics. The correlation integral $C(r)$ for the strain rate $5.66 \times 10^{-4}$ sec$^{-1}$ is shown in Fig. 4 for time delay $\tau = 4$ and embedding dimensions m = 9. The correlation dimension is determined from the slope of this curve. The inset in the Fig. 4 shows the variation of the correlation dimension with the embedding dimension. The convergent value of the slope so obtained gives a correlation dimension of 6.5±0.2. The value of the correlation entropy ($K_2$) obtained for the entire strain rate regime is ~ 0.03. This supports the presence of deterministic chaos in the PLC effect of the low carbon steel.

S. J. Noronha et. al. [9] have showed the existence of chaos in the PLC effect of the substitutional Al-Mg alloy. They obtained the maximum Lyapunov exponent to be around 1, which is much higher than that of the low carbon steel. The correlation dimension for the Al-Mg alloy was 3.2 [9], which is much smaller than that of the low carbon steel. The correlation dimension provides information on the minimum number of dynamical variables required to model a system and is often considered as a measure of its complexity [27]. Hence, the number of degrees of freedom in the PLC effect of the low carbon steel is more compared to the Al-Mg alloy. This also indicates the higher complexity and multiple deviations of the dynamics of the PLC effect in the interstitial alloy as compared to the substitutional alloy.

To compare the complexity of the PLC effect in the interstitial and substitutional alloy, we performed the Multiscale Entropy (MSE) analysis [28] of the PLC data from low carbon steel and Al-2.5%Mg alloy. For identical strain rate it is seen that the Sample Entropy of the low carbon steel is more than that of the Al-2.5%Mg alloy. A typical result of the MSE analysis is shown in Fig. 5 for the strain rate of $1.95 \times 10^{-4} \, s^{-1}$. This supports the fact that the PLC dynamics is more complex in case of the interstitial alloy.

A possible approach for understanding the observed complexity measure is to consider the dislocation-solute interaction mechanisms in these two alloys. By acting as a source of strain field, dislocations interact with solute atoms in metallic alloys. The origin of the dislocation-solute interaction in metallic alloys can be divided into two major effects, namely the size misfit effect and the modulus effect. The size misfit effect arises from the size difference between the host atom and the misfitting solute atom. This component of interaction is directly proportional to the size difference. The modulus interaction occurs if the presence of a solute atom locally alters the modulus of the crystal. The strain field generated is directly proportional to the difference in the elastic moduli of the host and the solute atom. But this interaction is only important when the other interactions are absent or negligible.

In case of the substitutional alloy Al-Mg, the size mismatch between the host (Al) and the solute (Mg) is high (about 12%) [29]. This spherically symmetric size misfit effect is the dominant part of the dislocation-solute interaction over the modulus effect [30]. Thus in the Al-Mg alloy, the solute atoms primarily interact with the edge dislocations. In contrary, due to the tetragonal lattice distortion caused by the interstitial carbon atom in low carbon steel, both edge and screw dislocations participate in the DSA of this alloy. It has been shown that the interaction between a screw dislocation and a carbon atom in iron has about the same strength as the interaction between an edge dislocation and a carbon atom. However, the accumulation of carbon atoms around a screw dislocation is almost twice as many as around an edge dislocation [31]. This reflects in the higher complexity of the PLC effect in the low carbon steel.

We also investigated the statistical properties of the stress-time series data for the two types of alloys. We have calculated the mean stress drop magnitude, the mean time required for the stress to increase (often referred to as the waiting time) and their corresponding standard deviations (std). The results evoke that the mean waiting time is of the same order in both the alloys. However, the mean stress drop magnitude for the low carbon steel is much smaller than that of the Al-Mg alloy. Moreover, the std of the stress drop magnitude is higher for the Al-Mg alloy.

The PLC instability is primarily governed by the cooperative movement of the dislocation forming the deformation band. The width of the band, in turn the number of dislocations in the band, primarily regulates the stress drop magnitude. The smaller stress drop magnitude for the low carbon steel is associated with the lesser number of dislocations participating in the band. This may be attributed to the fact that the stress field of the dislocation gets saturated [32] within a short time due to the combined effect of the larger diffusion coefficient and the rapid Snoek ordering of the interstitial carbon atoms. This forbids the bandwidth to increase further.

In case of the substitutional alloy, the locking of the dislocations is accomplished by the comparatively slower diffusion of the solute atoms to the band. The solute diffusion in this case cannot saturate the stress field of the dislocations. This allows the bandwidth to increase during the waiting time, resulting in larger stress drops. Moreover, the diffusion of solute atoms in this case is vacancy assisted. These vacancies are produced during the deformation and their spatial distribution changes with strain. This induces variability in the solute diffusion mechanism in case of the Al-Mg alloy and is reflected as the high dispersion of the stress drop magnitude.

In conclusion, we showed that the PLC effect in the low carbon steel alloy exhibits marginal deterministic chaotic dynamics in the studied strain rate regime. Moreover, a comparative study of the chaotic behavior of the PLC effect in this interstitial alloy and the substitutional Al-Mg alloy showed that the PLC dynamics in the interstitial alloy is more complex compared to the substitutional alloy.

**Figure captions:**

Fig.1. The segment of the drift corrected true stress fluctuation vs. time curve for the strain rate $5.66 \times 10^{-4}$ sec$^{-1}$.

Fig.2. Singular value spectrum of the stress time series data for the strain rate $5.66 \times 10^{-4}$ sec$^{-1}$. $s_i$'s are the singular values arranged in descending order.

Fig.3. Projection of the reconstructed attractor onto two-dimensional subspace spanned by the first two eigen vectors $C_1$ and $C_2$ for the strain rate $5.66 \times 10^{-4}$ sec$^{-1}$.

Fig.4. The Variation of the Correlation integral $C(r)$ with $r$ in for the embedding dimension m = 9. The inset shows the variation of the Correlation Dimension ($v$) with the embedding dimension ($m$).

Fig.5. The MSE analysis of the PLC effect in the low carbon steel and Al-2.5%Mg alloy deformed at a strain rate $1.95 \times 10^{-4}$ sec$^{-1}$.

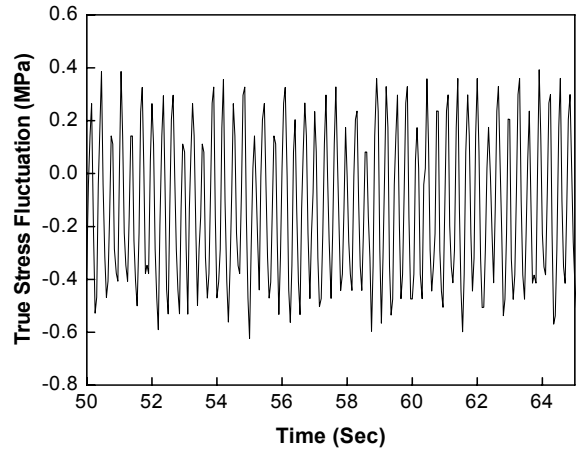

Fig. 1

.

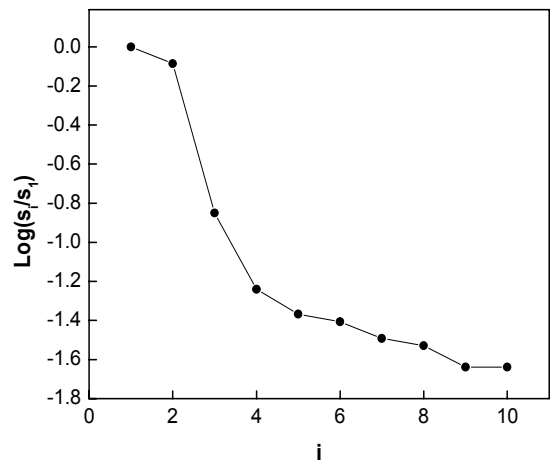

Fig. 2

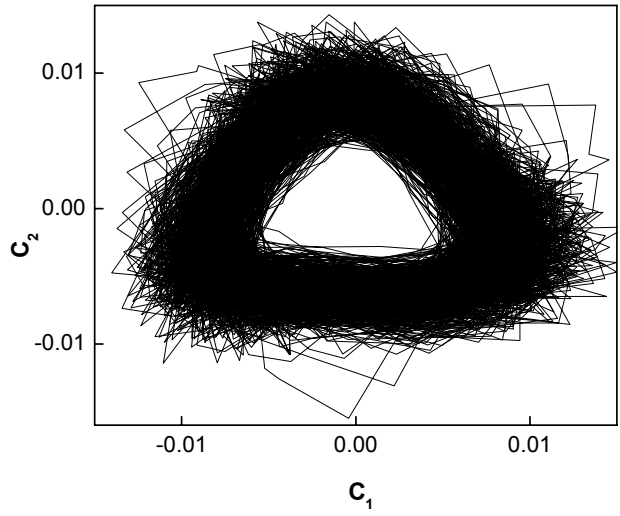

Fig. 3

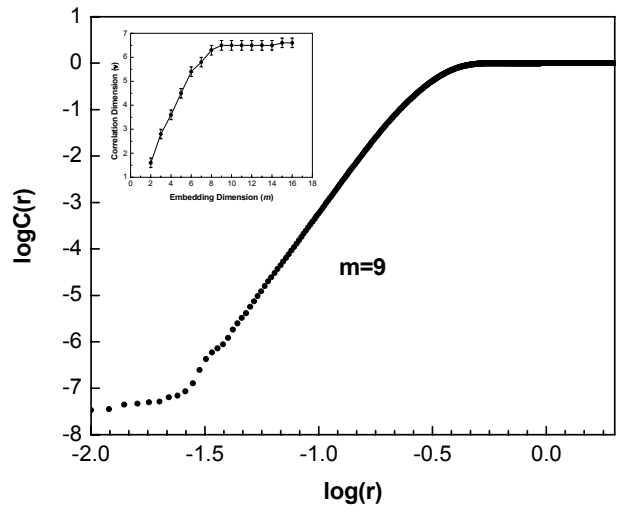

Fig. 4

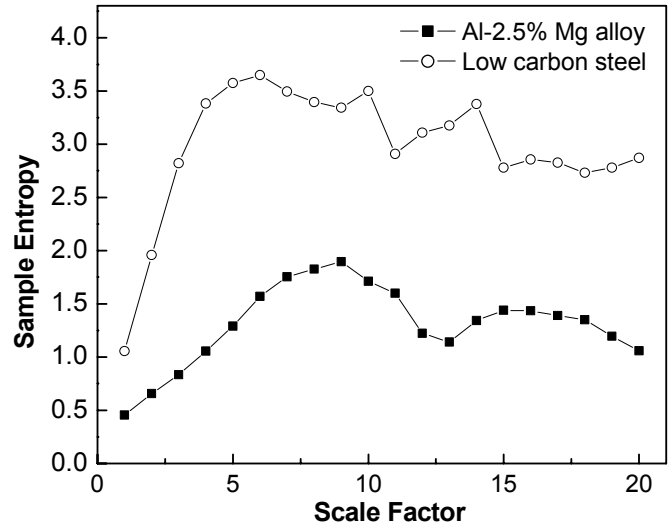

Fig. 5